%% file: main.tex
\documentclass[acmsmall,screen,nonacm,prologue,table]{acmart}

\AtBeginDocument{%
  }

\setcopyright{acmlicensed}
\copyrightyear{2024}
\acmYear{2024}
\acmDOI{XXXXXXX.XXXXXXX}

\newif\ifdraft
\draftfalse

\usepackage{subcaption}
\usepackage{soul}
\renewcommand{\rq}[1]{RQ\textsubscript{#1}}
\usepackage{booktabs}
\usepackage{arydshln}
\ifdraft
    \usepackage[draft,commandnameprefix=ifneeded]{changes}
\else
    \usepackage[final,commandnameprefix=ifneeded]{changes}
\fi
\usepackage{algorithmicx}
\usepackage{algorithm}
\usepackage{algpseudocode}
\usepackage{multirow}

\newcommand{\multrow}[1]{\begin{tabular}{@{}c@{}} #1 \end{tabular}}

\newcommand{\cachealot}{Cache-a-lot}

\begin{document}

\input{content/header}

\input{content/abstract}

% \received{12 September 2024}
% \received[revised]{DD M YYYY}
% \received[accepted]{DD M YYYY}

\maketitle

% \ifdraft
% TODOs:
% \begin{itemize}
%     \item threads to validity??
% \end{itemize}

% Done:
% \begin{itemize}
%     \item Memory consumption: make a separate RQ1.5 for it
%     \begin{itemize}
%         \item "static" memory consumption: measure the increase in memory after running all benchmarks
%         \item peak memory consumption: measure time with different max heap values
%         \item also add actual peak consumption experiments?
%         \item optimizations: add info about peak consumption on a formula
%     \end{itemize}

%     \item fix all typos

%     \item save the current version

%     \item Hash function
%     \begin{itemize}
%         \item pseudocode?
%         \item explicitly state that Candidate Testing section resolves hash collisions
%     \end{itemize}

%     \item Missing discussion of incremental solving and knowledge compilation in related work
%     \begin{itemize}
%         \item add that we are focused on external caches
%         \item add words about " we are focused on regular solving, but our approach can be adapted for incremental solving as well"
%         \item read about knowledge compilation and add additional paragraph into related work section
%     \end{itemize}

%     \item definition of partial substitution through exists

%     \item fix presentations of tables~\ref{tab:benchmarks} and~\ref{tab:t1_rq1}

%     \item add info about caching methodology setup (questions B.3-4)
% \end{itemize}
% \fi

\input{content/introduction}

\input{content/background}

\input{content/related}

\input{content/approach}

\input{content/implementation}

\input{content/evaluation}

\added{\protect\input{content/new/discussion}}
% \vspace*{-0.5em}
\input{content/conclusion}

% \begin{acks}
% \end{acks}

\bibliographystyle{ACM-Reference-Format}
\bibliography{main}

\end{document}

%% file: content/header.tex
\title{\cachealot: Pushing the Limits of Unsatisfiable Core Reuse in SMT-Based Program Analysis}

\author{Rustam Sadykov}
\email{rustam.sadykov@jetbrains.com}
\orcid{...}
\affiliation{%
  \institution{JetBrains Research}
  \city{Munich}
  \country{Germany}
}

\author{Azat Abdullin}
\orcid{...}
\affiliation{%
  \institution{JetBrains Research}
  \city{Amsterdam}
  \country{The Netherlands}}
\email{azat.abdullin@jetbrains.com}

\author{Marat Akhin}
\orcid{...}
\affiliation{%
  \institution{JetBrains Research}
  \city{Amsterdam}
  \country{The Netherlands}
}
\email{marat.akhin@jetbrains.com}

%% file: content/abstract.tex
\begin{abstract}

Satisfiability Modulo Theories~(SMT) solvers are integral to program analysis techniques like concolic and symbolic execution, where they help assess the satisfiability of logical formulae to explore execution paths of the program under test. However, frequent solver invocations are still the main performance bottleneck of these techniques. One way to mitigate this challenge is through optimizations such as caching and reusing solver results. While current methods typically focus on reusing results from fully equivalent or closely related formulas, they often miss broader opportunities for reuse.

In this paper, we propose a novel approach, \cachealot{}, that extends the reuse of unsatisfiable~(unsat) results by systematically considering all possible variable substitutions. This enables more extensive reuse of results, thereby reducing the number of SMT solver invocations and improving the overall efficiency of concolic and symbolic execution.

Our evaluation, conducted against the state-of-the-art Utopia solution using two benchmark sets, shows significant improvements, particularly with more complex formulas. Our method achieves up to 74\% unsat core reuse, compared to Utopia's 41\%, and significant increase in the time savings. These results demonstrate that, despite the additional computational complexity, the broader reuse of unsat results significantly enhances performance, offering valuable advancements for formal verification and program analysis.

\end{abstract}


\ccsdesc[500]{Software and its engineering~Software defect analysis}
\ccsdesc[500]{Software and its engineering~Formal software verification}
\ccsdesc[500]{Theory of computation~Logic and verification}
\ccsdesc[500]{Computing methodologies~Symbolic and algebraic manipulation}

\keywords{Symbolic program analysis, symbolic execution, concolic execution, SMT solver, solution reuse, caching}

%% file: content/introduction.tex
\section{Introduction}
\label{section:intro}

Satisfiability Modulo Theories~(SMT)~\cite{SMT} solvers are integral to various applications, including formal verification~\cite{SMTForVerification}, automated reasoning~\cite{SMTForReasoning}, and program analysis. In program analysis techniques like concolic~\cite{ConcolicTesting} and symbolic~\cite{SymbolicExecution} execution, SMT solvers are essential for exploring different execution paths by assessing the satisfiability of logical formulae. However, the computational cost of these solver invocations can be substantial, prompting the development of numerous optimisations to improve the efficiency of the analysis techniques.

One key optimisation technique that proved its effectiveness throughout the years is caching and reusing of SMT solver results to minimize the number of solver calls. This technique relies on the fact that formulae, generated during program analysis, often exhibit \deleted{significant} similarity, allowing for the reuse of previously computed results to avoid redundant computations and enhance performance.

\added{SMT caching strategies can be divided into two main categories: internal and external. External cache is placed outside the SMT solver, effectively treating the solver as a black box. This design choice allows the caching layer to operate independently of the solver's internals, ensuring compatibility with any SMT solver without the need for modifications. In contrast, an internal cache is optimized for a particular SMT solver, leveraging solver-specific internal knowledge to achieve greater efficiency. This work focuses on external caching techniques.}

% \added{Caching strategies come in various forms. One common type is external caching, where the caching mechanism is placed outside the SMT solver, effectively treating the solver as a black box. This design choice allows the caching layer to operate independently of the solver's internals, ensuring compatibility with any SMT solver without the need for modifications.}

\added{External} SMT caching techniques \replaced{can be further categorised}{can be divided into two main categories} depending on the type of SMT solver result that they rely on: satisfiable~(sat) or unsatisfiable~(unsat) result reuse. These two categories use different approaches and \added{are} usually independent from each other. Therefore, almost any existing sat reuse approach can be combined with any unsat reuse approach.

In this paper, we focus on improving the reuse of unsat results. Existing methods typically enable result sharing only between fully equivalent or directly related formulae. While effective, these methods often miss opportunities for more extensive result reuse. Additionally, existing methods often rely on some form of formulae preprocessing, mainly formula canonization. 

Utopia~\cite{Utopia} is the state-of-the-art approach for SMT caching. The original Utopia paper incorporates both sat and unsat reuse approaches~(that are independent of each other), and in this paper, we will use ``Utopia'' as ``Utopia unsat reuse approach''. Utopia introduced the idea of hashing and filtering of SMT results, which allows it to \deleted{efficiently} detect similarities between SMT formulae and reuse unsats. However, Utopia is still subject to aforementioned problems: it only considers fully equivalent or directly related formulae and \deleted{it} heavily relies on the formula canonization.

We present \cachealot{}, a novel Utopia-based approach that addresses these limitations by efficiently considering all possible variable substitutions, facilitating a broader reuse of unsat results. This advancement allows it to significantly reduce the number of SMT solver invocations, thereby enhancing the efficiency of concolic and symbolic execution. Additionally, \cachealot{} does not rely on the formulae preprocessing, making it more flexible than existing approaches.

We evaluated our approach against Utopia on two \deleted{sets of} benchmarks: an original Utopia benchmark containing simpler formulae and a new benchmark with more complex formulae. Our approach demonstrated significant improvements in both cases. While gains were modest for simpler formulae, \added{Cache-a-lot's 34.47\% of unsat reuse ratio compared to Utopia's 33.30\%,} our method showed substantial benefits for complex formulae, \added{74.36\% compared to 40.50\%.} \deleted{notably in reducing the number of solver invocations and overall solving time.} Furthermore, we found that Utopia's performance can vary significantly with different canonization strategies, whereas our method systematically evaluates all possible renamings, avoiding such inconsistencies. Although our approach involves more complex computations, the enhanced result reuse and reduced solver invocations provide substantial overall time savings, justifying the additional computational effort.

This paper is organized to provide a comprehensive overview of our approach and its implications. Section~\ref{section:background} lays the groundwork by discussing the motivation behind our work, explaining relevant terminology, and addressing the problem we aim to solve. We also present properties and observations on which existing solutions are based and introduce the key observation that is the ground of our approach. Section~\ref{section:relatedWork} focuses on related work, reviewing the current state of research to highlight how our approach fits within and extends existing knowledge. Section~\ref{section:approach} details our proposed approach, explaining how we identify candidate formulae, determine the necessary substitutions, and implement optimizations to enhance efficiency. Section~\ref{section:impl} provides an overview of the implementation of our approach, addressing the key aspects and practical considerations involved in implementing our approach. In Section~\ref{section:evaluation}, we present an evaluation of our approach, including benchmarks, research questions, and experimental results. This section provides a qualitative and quantitative analysis of our approach’s efficiency and discusses the outcomes of described optimizations. \added{Section~\ref{section:discussion} discusses the broader implications of our approach and its relationship to related techniques, including incremental solving, knowledge compilation, and graph-based methods. We explore possible extensions to Cache-a-lot and analyze the trade-offs associated with integrating these techniques.} Finally, Section~\ref{section:conclusion} wraps up the paper by summarizing the key findings from our evaluation, discussing their implications, and suggesting potential directions for future research.

%% file: content/background.tex
\section{Reusing Unsatisfiable Cores of Formulae}
\label{section:background}

In this paper, we propose an innovative approach to enhance the efficiency of SMT solvers, particularly during symbolic and concolic analysis. The core idea is to store previously encountered formulae alongside their corresponding SMT solver results and reuse them to determine the satisfiability of new formulae, avoiding unnecessary SMT solver invocations. Unlike most of the existing state-of-the-art techniques, our approach is theory-independent. It extends the scope of result reuse by incorporating a broader set of equivalence and implication relations, including those that involve variable substitutions.

\subsection{Motivating Example}

Consider a scenario where two formulae are encountered during a computational process:
\begin{equation}
    (x > y) \land (y > z) \land (z > x)
    \label{eq:firstExample}
\end{equation}
\begin{equation}
    (b > c) \land (c > d) \land (d > b) \land (a > b)
    \label{eq:secondExample}
\end{equation}

We aim to solve these formulae using an SMT solver. When invoked on each formula separately, the solver determines that both are unsatisfiable. Additionally, SMT solvers can generate an unsatisfiable core (unsat core) for each formula, a subset of the formula's conjuncts~(clauses) that is itself unsatisfiable~\cite{UnsatCore}. It should be noted that any formula can potentially have multiple unsat cores, and a solver can return any one of them.

Formula~\ref{eq:firstExample} has a single unsat core $[x > y, y > z,z > x]$ and 
formula~\ref{eq:secondExample} has a single unsat core $[b > c, c > d, d > b]$. Despite the
fact that these two formulae do not share any common clauses, state-of-the-art approaches
like Utopia are able to reuse the results of one formula to determine the unsatisfiability
of another. That works only because existing approaches rely on formula canonization: all the variables are renamed into some canonical form in order of their appearance. However, this
approach has one major disadvantage: the efficiency of caching is heavily reliant on the order
of clauses~(and variables) inside a formula. For example, formula~\ref{eq:thirdExample} is
logically equivalent to~\ref{eq:secondExample}, but Utopia is not able to reuse the results of one formula to solve another, necessitating separate SMT solver invocations.
\begin{equation}
    (a > b) \land (b > c) \land (c > d) \land (d > b)
    \label{eq:thirdExample}
\end{equation}

Still, it is obvious to a human observer that the clauses of formula~\ref{eq:firstExample}
are structurally equal to the clauses of unsat core of formula~\ref{eq:thirdExample}, meaning that they only differ in the variable names. Substitution $\{x \mapsto b, y \mapsto c, z \mapsto d\}$  transforms one set of clauses into another, thus proving the structural equality. This observation implies that by considering variable substitutions, we can determine that the \replaced{third}{second} formula is also unsatisfiable without additional invocation of the SMT solver. In other words, the solver result from the first formula can be reused to establish the unsatisfiability of the \replaced{third}{second}, thereby reducing computational overhead.

This example illustrates how extending the caching mechanisms of SMT solvers to consider variable substitutions and making them independent of the canonization/clause order can significantly enhance their efficiency.

\subsection{Main Properties}

Now, we formalize the observations made in the motivating example. Current methods for caching unsatisfiable results are based on the following key properties of SMT formulae:

\begin{itemize} 
    \item Two equivalent formulae share their satisfiability. 
    \item If a formula implies an unsatisfiable formula, the original formula is also unsatisfiable. 
    \item An unsat core, by definition, is unsatisfiable. 
\end{itemize}

As demonstrated in our motivating example, the effectiveness of caching techniques depends on the accuracy of identifying equivalence and/or implication relations between formulae. Existing approaches, while utilizing formula canonization to rename variables uniformly, do not fully account for cases involving variable substitutions.

To address this gap, we introduce an additional property: a formula obtained through a variable substitution on an unsatisfiable formula remains unsatisfiable. This principle allows us to generalize unsat core results across a broader range of formulae. In practice, this means that if we can identify a substitution for an unsat core such that a formula subsumes the substituted unsat core, we can state that the formula is unsatisfiable without further SMT solver invocation.

By developing a new approach for unsat core caching that considers variable substitution, we can significantly reduce the number of redundant SMT solver invocations, leading to more efficient analysis overall. However, determining the appropriate substitution efficiently is not a trivial task. Brute force implementation will fail due to time/memory limitations,
therefore, an important part of our work is an optimisation of variable substitution search.

%% file: content/related.tex
\section{Related Work}
\label{section:relatedWork}

In recent years, several approaches have been proposed to improve the efficiency of SMT solvers by caching and reusing previously computed results. These approaches vary in their algorithms, target logic, and effectiveness. Below, we discuss some of the most prominent methods, such as Klee~\cite{Klee}, Green~\cite{Green}, Recall~\cite{Recal}, GreenTrie~\cite{GreenTrie}, and Utopia~\cite{Utopia}.

Klee~\cite{Klee} is a symbolic execution tool that includes two distinct caching frameworks: the branch cache and the counterexample cache. These caching mechanisms are specifically designed to handle formulae in the quantifier-free theory of bit-vectors and bit-vector arrays~(QF\_ABV). The branch cache memorizes formulae and their corresponding solutions, allowing quick retrieval when the same formula is encountered again. On the other hand, the counterexample cache works by determining whether a target formula is a subset or a superset of previously solved formulae. If the target formula is contained within a satisfiable or unsatisfiable formula, Klee can \replaced{quickly}{immediately} conclude its satisfiability or unsatisfiability without additional solver invocations.

Green~\cite{Green} is a caching framework that targets formulae within quantifier-free linear integer arithmetic logic~(QF\_LIA). Unlike Klee, Green applies a set of predefined simplification rules to the target formula before checking its satisfiability. If the simplified formula matches a formula in the database that has already been solved, Green can determine the satisfiability of the target formula based on the stored result. This approach effectively reduces the need to resolve the same or similar formulae, thus improving the efficiency of the solver. 

Recal~\cite{Recal} is another caching approach that targets QF\_LIA formulae. Similar to Green, it applies a set of predefined simplification rules to the formula and produces canonical matrix representation. This matrix representation can be later used to efficiently check if the formula is contained in the database of the previously encountered formulae.

GreenTrie~\cite{GreenTrie} builds upon the Green framework by introducing the ability to identify logical implications between formulae. Specifically, GreenTrie checks if a target formula implies or is implied by previously solved formulae. If the target formula is implied by a satisfiable formula, GreenTrie concludes that the target is also satisfiable. Conversely, if the target formula implies an unsatisfiable formula, it is deemed unsatisfiable. This extension enhances Green's ability to reuse previously computed results, particularly in scenarios with logical implications.

Utopia~\cite{Utopia} represents a significant advancement in caching frameworks, offering distinct methodologies for reusing solutions to satisfiable and unsatisfiable formulae.
For satisfiability caching, Utopia employs a heuristic called Sat-delta. This heuristic identifies formulae that may share solutions with a given target formula, even when the formulae are not structurally identical. This allows Utopia to extend reusability beyond strictly equivalent or syntactically similar formulae. However, this flexibility comes with a limitation: the Sat-delta heuristic is confined to quantifier-free logic, which restricts its applicability to certain types of formulae.

On the other hand, Utopia's approach to unsatisfiability caching centres around Unsat-footprint heuristic. Unlike the Sat-delta heuristic, Unsat-footprint is not limited by any specific logic, making it applicable across a broader range of formulae. This unsat core reuse methodology allows Utopia to identify and leverage previously computed unsat cores effectively. However, while the Unsat-footprint heuristic is logic-independent, its effectiveness depends on how the unsat cores relate to the target formula.

Combining these heuristics --- Sat-delta for satisfiability and Unsat-footprint for unsatisfiability --- Utopia offers a versatile and efficient caching framework. It strikes a balance between logic-specific limitations in satisfiability caching and broader applicability in unsatisfiability caching, making it one of the most advanced tools in the field.

Although each approach has made progress in reusing unsatisfiable formulae, there are still some limitations. Klee, Green, Recal and GreenTrie primarily focus on satisfiability caching and rely on specific logic, which restricts their general applicability. Utopia takes a broader approach with its Unsat-footprint heuristic, enabling the efficient reuse of unsat cores across different theories without depending on a particular logic. Despite these advancements, none of the existing approaches fully exploit the potential of variable substitution in unsat core reuse. They only rely on formula canonization, which further adds
an additional drawback of being reliant on the order of clauses inside the formula.
Our idea is to address both of these gaps by considering variable substitution within the caching mechanism, allowing for more flexible and efficient reuse of unsatisfiable results.

%% file: content/approach.tex
\section{\cachealot{} Approach}
\label{section:approach}

\begin{figure}[tbh]
    \centering
    \includegraphics[width=0.75\linewidth]{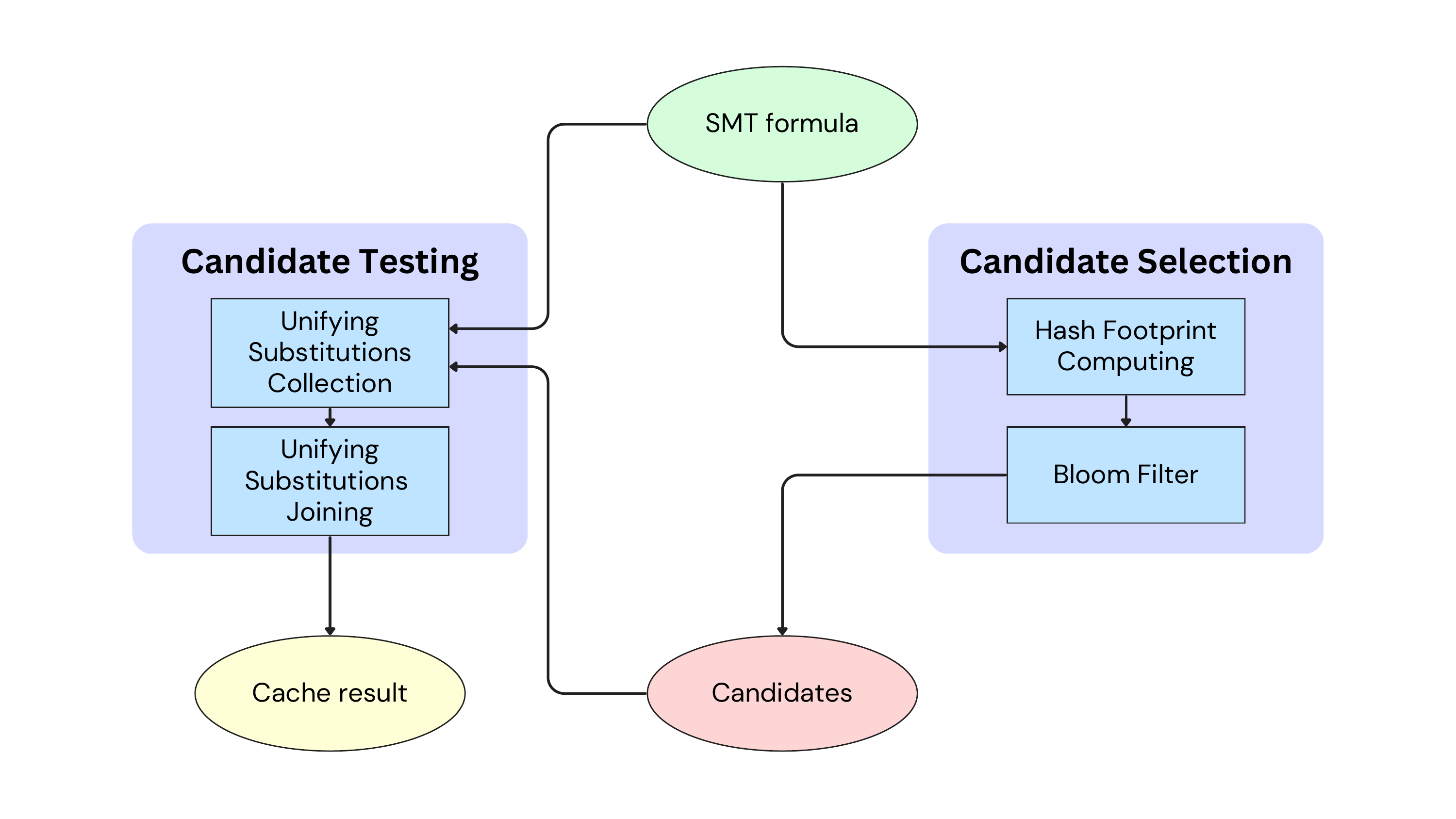}
    \caption{Pipeline of the SMT Formula Reuse Process: The flowchart illustrates the steps involved in reusing unsat cores, beginning with candidate selection using a Bloom filter and proceeding to candidate testing by collection and joining unifying variable substitutions.}
    \label{fig:flow}
\end{figure}

\cachealot{} approach is designed to optimize the reuse of SMT formulae solutions by focusing on the efficient storage, search, and testing of unsat core candidates. The approach is divided into two main components:
\begin{itemize}
    \item \textbf{Candidate Selection}: given a formula $F$, select a set of unsat core candidates $S$, where each candidate can be potentially used to check the unsatisfiability of $F$. An unsat core is selected as a candidate $s \in S$ when each clause of the unsat core is assumed to be structurally present in the formula $F$.
    \item \textbf{Candidate Testing}: given a formula $F$ and a set of
    candidates $S$, check if there exists a candidate $s \in S$
    that implies the unsatisfiability of $F$.
\end{itemize}
 
 We will begin by exploring these components. Finally, we will discuss various optimisations that further enhance the efficiency and effectiveness of the \cachealot{} approach. The overall pipeline is depicted in the figure~\ref{fig:flow}.

\subsection{Candidate Selection}

This component follows the Utopia~\cite{Utopia} approach but incorporates a modified hashing technique explicitly tailored to align with the \cachealot{} framework. The primary objective is to efficiently and swiftly match each given formula $F$ to a set of unsat core candidates $S$, selected from the set of all unsat cores detected during the analysis.

This step consists of two main stages:
\begin{itemize}
    \item Computing a unique hash footprint for $F$.
    \item Selecting all unsat core candidates for $F$ that match its hash footprint.
\end{itemize}

\subsubsection{Hash Footprint}
\hfill

Hash footprint of a formula $F$ is a list of hashes $[h_1, h_2, ..., h_n]$, where \added{$F$ is in conjunctive form} and $h_i$ is a hash of the i-th clause of $F$. The idea behind the hash footprint is to provide a simplified representation of a formula's clauses. It is assumed that the hash footprint closely reflects the structure of the formula's clauses to enable a faster unsat core candidate testing process.

\protect\input{content/old/hash}

\added{\protect\input{content/new/hash}}

\subsubsection{Bloom Filter}
\hfill

Given a formula $F$ and an unsat core $s$, we need to determine if $s$ is a valid unsat core candidate, i.e. whether clauses of $s$ are structurally present in $F$. This can be easily done using hash footprints:~if the hash footprint of $s$ is a subset of the hash footprint of $F$, then all the clauses of $s$ are structurally present in $F$. 

However, directly checking the subset relation between hash footprints can be computationally expensive, especially as their sizes increase. We use Bloom filter to address this problem. We represent each hash footprint as a bitset with the fixed size $N$. Each hash from the hash footprint is used as an index~(modulo $N$) in the bitset to set the value of the corresponding bit to 1.

Once the hash footprints are represented as bitsets, checking for a subset relation becomes significantly faster. This is because the subset check can be reduced to a simple bitwise operation between the bitsets, which is much more efficient than comparing individual elements of the hash footprints. This use of Bloom filters thus enhances the performance of the \cachealot{} approach by enabling rapid pre-filtering of potential matches before more detailed analysis is conducted. As the result of the candidate selection stage, for each formula $F$, we get a set of unsat core candidates~$S$.

\subsection{Candidate Testing} \label{sec:testing}

In this part, we will explain how to test candidates efficiently. We will begin by outlining the idea behind the testing process. Following that, we will delve into each specific aspect of the testing procedure, providing a detailed explanation of each step. Finally, we will conclude with a description of the optimisations that enhance the overall efficiency of the testing process.

\subsubsection{The Idea}
\hfill

The main idea behind testing an unsat core candidate $s$ against a formula $F$ is to determine whether a substitution $\sigma$ exists for the candidate such that the formula subsumes it: $\sigma(s) \subseteq F$. This approach expands the potential for reusing unsat cores beyond what is achievable with state-of-the-art methods. However, directly iterating over all possible substitutions would be extremely slow, thereby necessitating process optimisation.

To illustrate the main concept, let us consider an example with an unsat core candidate consisting of clauses $[x > y,\, y > z,\, z > x]$ and a formula consisting of clauses $[a > b,\, b > c,\, c > d,\, d > b]$. The first step involves calculating unifying substitutions for each pair of clauses independently, as shown in the table~\ref{tab:substitutions}. A unifying substitution is a mapping that maps variables from one expression to corresponding variables in another expression, making the two expressions identical~\cite{UnifyingSubstitution}. In the context of SMT formulae, the existence of unifying substitution implies that two clauses are structurally equivalent. This equivalence is crucial for matching an unsat core clause with a formula clause. By finding a consistent way to substitute variables across different clauses, it becomes possible to determine if a formula can subsume an unsat core candidate.

In the context of our work, we will introduce the following definitions:
\begin{itemize}
    \item \textbf{Unifying substitution} --- a mapping $\sigma$ that replaces variables of one clause $c_1$ with the corresponding variables from another clause $c_2$ to make the two clauses identical: $\sigma(c_1)~\equiv~c_2$.

    \item \textbf{Complete substitution} --- a mapping $\sigma$ that replaces all variables of an unsat core candidate $s$ with the corresponding variables from the formula $F$ so that the formula subsumes the candidate: $\sigma(s) \subseteq F$.

    \item \textbf{Partial substitution} --- \replaced{a mapping $\sigma$ that replaces all variables of a subset $s'$ of an unsat core candidate $s$ with the corresponding variables from the formula $F$ so that the formula subsumes the subset of the candidate: $\exists s' \subseteq s$ such that $\sigma(s') \subseteq F$.}{a mapping $\sigma$ that replaces some of the variables of an unsat core candidate $s$ with the corresponding variables from the formula $F$ so that the formula subsumes some of the clauses $[c_i, ..., c_j]$ of the candidate: $\forall k, i \leq k \leq j, \sigma(c_k) \in F$.}

\end{itemize}

\begin{table}[tbh]
    \begin{tabular}{|c|c|c|c|c|}
        \hline
        & $a > b$ & $b > c$ & $c > d$ & $d > b$\\
        \hline
        $x > y$ & $\{x \mapsto a, y\mapsto b\}$ & \cellcolor{green!25}$\{x \mapsto b, y\mapsto c\}$ & $\{x \mapsto c, y\mapsto d\}$ & $\{x \mapsto d, y\mapsto b\}$\\
        \hline
        $y > z$ & $\{y \mapsto a, z\mapsto b\}$ & $\{y \mapsto b, z\mapsto c\}$ & \cellcolor{green!25}$\{y \mapsto c, z\mapsto d\}$ & $\{y \mapsto d, z\mapsto b\}$\\
        \hline
        $z > x$ & $\{x \mapsto b, z\mapsto a\}$ & $\{x \mapsto c, z\mapsto b\}$ & $\{x \mapsto d, z\mapsto c\}$ & \cellcolor{green!25}$\{x \mapsto b, z\mapsto d\}$\\
        \hline
    \end{tabular}
    \caption{Unifying substitutions for each pair of clauses}
    \label{tab:substitutions}
\end{table}

If a complete substitution for the unsat core candidate exists, like $\{x\mapsto b,\,y\mapsto c,\,z\mapsto d\}$ in this example, it means that each clause from the unsat core candidate matches a clause from the formula with this substitution. The unifying substitutions highlighted in the table~\ref{tab:substitutions} align with the complete substitution and can be combined to form it.

On the other hand, if no complete substitution exists, it implies that there is no way to match all clauses of the unsat core candidate with those in the formula using any substitution. This implies that there is no way to combine any consistent renaming from the unifying substitutions. In such cases, the unsat core candidate cannot be used to derive the unsatisfiability of the formula.

Rather than thoroughly checking every possible complete substitution, we can search for a sequence of consistent unifying substitutions that can be combined into one. \replaced{Although it might seem slow, it's actually faster because it avoids checking every clause with every substitution. This is especially important when working with large formulas.}{Although this might seem slow initially, it is much faster in practice because it reduces the need to repeatedly traverse and compare clauses with every iterated substitution, which is crucial when dealing with large formulae.}

In summary, the testing process consists of two key steps:
\begin{itemize}
    \item {\bf Substitution Collection}: Gather all possible unifying substitutions for each pair of clauses from the unsat core candidate and the formula.
    \item {\bf Substitution Joining}: Combine these unifying substitutions to construct a complete substitution for the unsat core candidate or determine that no such substitution exists.
\end{itemize}

\subsubsection{Substitution Collection}
\hfill

First, we need to find a unifying substitution for each pair of clauses from the unsat core candidate and the formula. This process, known as unification~\cite{Unification}, involves matching the structure of the clauses while accounting for differences in their variables.

To achieve this, we traverse the pairs of clauses, comparing each pair of nodes based on their type and number of arguments while building the required substitution. We maintain and update the current substitution throughout the traversal, ensuring consistency as we proceed. Several exceptional cases require careful handling:
\begin{itemize}
    \item {\bf Constants}: Their values must be identical.
    \item {\bf Quantifiers and Lambdas}: When dealing with quantifiers or lambda expressions, we must consider their bound variables and check the bodies of these expressions with an updated substitution. For instance, if we encounter a quantifier $\forall x$ in one clause and $\forall y$ in the corresponding clause, we temporarily substitute the bound variable $x$ with $y$, even when $x$ was already in the substitution. This action also helps handle shadowing, where an inner scope variable may hide an outer scope variable with the same name. As we traverse the bodies of these quantifiers, we treat $x$ and $y$ like regular variables, considering modified substitution. Once we finish traversing the bodies, we revert the substitution for bound variables to their previous states.
    \item {\bf Variables}: For example, if we encounter variables $a$ and $b$, and \replaced{$a$}{a} has not been substituted yet, we can update the substitution for $a$ with $b$. However, if $a$ already has a substitution, we must check if it matches $b$. If it doesn't, no valid substitution exists for this pair of clauses.
\end{itemize}

\subsubsection{Substitution Joining}
\hfill

Once we have collected the unifying substitutions, the next step is to construct a complete substitution for the unsat core candidate by combining these unifying substitutions. Let us refer back to the earlier example with the unsat core candidate $[x > y,\, y > z,\, z > x]$, the formula $[a > b,\, b > c,\, c > d,\, d > b]$ and corresponding unifying substitutions shown in the table~\ref{tab:substitutions}.

The complete substitution is constructed by selecting one of the unifying substitutions
from each row of the table~\ref{tab:substitutions} and combining them together. However,
while combining, we must also ensure consistency in variable mappings; e.g. if a
variable $x$ is mapped to $y$ in one of the unifying substitutions, \replaced{we must ensure that other unifying substitutions that contain $x$ also map it to $y$}{we must ensure that every other unifying substitution also maps $x$ to $y$~(if it contains it)}.

To understand this process better, we can think of the set of unifying substitutions for each clause~(i.e. each row of the table~\ref{tab:substitutions}) in the unsat core candidate as a table. In these table, the columns represent the variables in the original clause, and the rows represent the possible substitutions for these variables. For example, a set of unifying substitutions for the clause $x > y$ can be represented as a table presented in the figure~\ref{fig:unifyingSubstitutionsTable}.

\begin{figure}[tbh]
\begin{center}
    $\displaystyle\left\{ \begin{matrix}
            \{x \mapsto a, y\mapsto b\},\\
            \{x \mapsto b, y\mapsto c\},\\
            \{x \mapsto c, y\mapsto d\},\\
            \{x \mapsto d, y\mapsto b\}
        \end{matrix} \right\} \Longrightarrow$
    \begin{tabular}{|c|c|}
        \hline
        x & y\\
        \hline
        a & b\\
        b & c\\
        c & d\\
        d & b\\
        \hline
    \end{tabular}
\end{center}
\caption{Table representation of the set of unifying substitutions}
\label{fig:unifyingSubstitutionsTable}
\end{figure}

Thus, the process of combining $n$ sets of individual substitutions into a single complete unsat core candidate substitution can be transformed into the process of joining $n$ tables together into one. The natural join operation~\cite{Join} exactly satisfies our needs by combining tables based on their common attributes --- shared variables. For example, a join operation between the tables for clauses $x > y$ and $y > z$ will produce a partial substitution as presented in the figure~\ref{fig:joinExample}.

\begin{figure}[tbh]
\begin{minipage}{0.35\linewidth}
\begin{center}
\begin{tabular}{|c|c|}
        \hline
        x & y\\
        \hline
        a & b\\
        b & c\\
        c & d\\
        d & b\\
        \hline
    \end{tabular}
$\bigotimes$
\begin{tabular}{|c|c|}
        \hline
        y & z\\
        \hline
        a & b\\
        b & c\\
        c & d\\
        d & b\\
        \hline
    \end{tabular}
$=$
\begin{tabular}{|c|c|c|}
        \hline
        x & y & z\\
        \hline
        a & b & c\\
        b & c & d\\
        c & d & b\\
        d & b & c\\
        \hline
    \end{tabular}
\end{center}
\caption{Join operation on two sets of unifying substitutions}
\label{fig:joinExample}
\end{minipage}\hspace{0.1\linewidth}
\begin{minipage}{0.4\linewidth}
\begin{center}
\begin{tabular}{|c|c|c|}
        \hline
        x & y & z\\
        \hline
        a & b & c\\
        b & c & d\\
        c & d & b\\
        d & b & c\\
        \hline
    \end{tabular}
$\bigotimes$
\begin{tabular}{|c|c|}
        \hline
        x & z\\
        \hline
        b & a\\
        c & b\\
        d & c\\
        b & d\\
        \hline
    \end{tabular}
$=$
\begin{tabular}{|c|c|c|}
        \hline
        x & y & z\\
        \hline
        b & c & d\\
        c & d & b\\
        d & b & c\\
        \hline
    \end{tabular}
\end{center}
\caption{Join operation between a partial and unifying substitutions}
\label{fig:joinExamplePartial}
\end{minipage}
\end{figure}

This resulting table gives us a partial substitution that makes the pair of clauses $x > y$ and $y > z$ subsumed by the original formula. To find the complete substitution for the entire unsat core candidate, we would continue by joining this partial substitution table with the table for the clause $z > x$~(figure~\ref{fig:joinExamplePartial}). The result of that operation contains all the substitutions that make the entire unsat core subsumed by the formula, meaning it contains all of the complete substitutions.

% \begin{figure}[tbh]
% \begin{center}
% \begin{tabular}{|c|c|c|}
%         \hline
%         x & y & z\\
%         \hline
%         a & b & c\\
%         b & c & d\\
%         c & d & b\\
%         d & b & c\\
%         \hline
%     \end{tabular}
% $\bigotimes$
% \begin{tabular}{|c|c|}
%         \hline
%         x & z\\
%         \hline
%         b & a\\
%         c & b\\
%         d & c\\
%         b & d\\
%         \hline
%     \end{tabular}
% $=$
% \begin{tabular}{|c|c|c|}
%         \hline
%         x & y & z\\
%         \hline
%         b & c & d\\
%         c & d & b\\
%         d & b & c\\
%         \hline
%     \end{tabular}
% \end{center}
% \caption{Join operation between a partial and unifying substitutions}
% \label{fig:joinExamplePartial}
% \end{figure}

By performing these join operations, we ensure that the resulting complete substitution is consistent across all clauses in the unsat core candidate. If the result of joining all unifying substitution tables is non-empty, it indicates that a complete substitution exists that allows the formula to subsume the unsat core candidate. If the resulting table contains more than one row, it means that there are several consistent complete substitutions and any one of them can be used.

Table representation for unifying substitutions allows us to greatly simplify the process of combining them, as well as allowing us to implement several optimisations that make the whole process even more efficient.

\subsubsection{Optimisations}
\hfill

Without optimisations, the testing process, particularly the join operation, can be extremely time- and memory-intensive. To improve efficiency, several key optimisations have been implemented. We are going to highlight three of the most crucial of them.

The first optimisation is straightforward but effective: grouping \deleted{the }formula's clauses by their hash codes. This enables efficient retrieval of clauses that are likely to match a candidate's clause, thereby reducing the number of unification\deleted{ processe}s needed when testing all candidates against a specific formula.

The next crucial optimisation is filtering out invalid unifying substitutions before the joining stage. After the substitution collection phase, we compute the set of possible substitutions $s_{x,i}$ separately for each variable $x$ within a clause $i$. Intersecting these sets across all clauses of the formula allows us to receive the set of valid substitutions for each variable $s_x$. Using these sets, we can filter out any unifying substitutions that map any variable $x$ to a value $y$ that is not contained in $s_x$. This process narrows down the potential substitutions to only those that could contribute to a consistent solution, thereby \deleted{significantly }reducing the computational workload before the join operation.

Finally, an essential optimisation is recognizing that we only need to find one complete substitution for the unsat core candidate rather than computing all possible complete substitutions. This insight leads to a more memory-efficient approach for handling the join operation. Instead of storing all potential combinations, we use an iterator that processes one clause at a time. Moving through the clause's table's rows, we check them against the current substitution. If a match is found, we continue to the following clause; if no matches are found, we revert to the previous clause to explore other options. This method reduces memory usage and avoids the need to generate and store large tables.

Additionally, general optimisations for table joining, such as table indexing, sorting tables by size, and using advanced iterators and join operation schedulers, further enhance the overall performance~\cite{Join}.

%% file: content/old/hash.tex
\deleted{Each clause of a formula is represented as an abstract syntax tree~(AST). The hashing process operates at each node of the AST by following a two-step approach:}

\deleted{$\bullet$ Computing a local hash code that reflects the specific entity represented by the node.}

\deleted{$\bullet$ Recursively combining this local hash code with those computed at the child nodes.}

\deleted{The local hash code of each node accounts for the node's sort, class, and content. Depending on the type of node, its contents are accounted for differently:}

\deleted{$\bullet$ {\bf Leaf nodes representing constants}: The local hash code is computed directly from the value of the node.}

\deleted{$\bullet$ {\bf Leaf nodes representing variables}: No hash code of the content is computed for these nodes to ensure that variable names are not considered, focusing on structural rather than nominal properties. This is the main difference from the original Utopia approach, as it heavily relies on the canonical variable names.}

\deleted{$\bullet$ {\bf Non-leaf nodes representing operators}: The hash code is influenced by the number of arguments~(e.g. number of bound variables in a quantifier) to capture the structural complexity of the operation.}

\deleted{Finally, hash code of each AST is computed by recursively combining local hash codes of each of its nodes using a hash combination algorithm. This recursive combination ensures that the hash value accurately reflects the entire structure of the formula clause. Thus, each hash in a hash footprint encapsulates the structural representation of the corresponding clause, and the hash footprint of a formula encapsulates structural properties of all of its clauses.}

%% file: content/new/hash.tex
Hash footprint computation is divided into two steps: 
\begin{itemize}
    \item calculating hash code for each clause of the formula;
    \item combining the hash codes of the formula's clauses into a hash footprint.
\end{itemize}

Let's take a look at each of these steps. Each clause of the formula is represented as an AST node. Algorithms~\ref{alg:hash_ast} presents how the hash code for an AST node is computed. The algorithm separately handles leaf~(line 3) and non-leaf~(line 9) nodes. Leaf nodes can be of two types: constants and variables. For constants hash code is defined as the combination of the constant's sort and it's value~(line 5). For variables the hash is computed from the node's sort, deliberately ignoring their name~(line 7). This ensures that the hash code captures the \emph{structure} of the formula and not the concrete variable positions. For non-leaf nodes, algorithm first combines the hash codes of the node's sort, type~(operator) and number of arguments~(line 10). Then, algorithm recursively calls $computeASTHash$ on node's children and combines their hashes with the node's hash~(line 14--15).

In our implementation, we use a standard Java $hashCode$ function to compute the initial hash values of the node attributes. Hash codes are combined together using $combineHashes$ function (see~\cite{HashCombine}) which fuses individual hashes into a single value, preserving the structure of the AST.

\ifdraft
\newenvironment{algocolor}{%
   \setlength{\parindent}{0pt}
   \itshape
   \color{blue}
\floatname{algorithm}{\color{blue}Algorithm}
}{}
\else
\newenvironment{algocolor}{}{}
\fi

\begin{algorithm}[tb]
\begin{algocolor}
\flushleft
\textbf{Input:} $node$ --- ASN node\\
\textbf{Output:} $localHash$ --- hash code of the given ASN node\\
\begin{algorithmic}[1]
\Function{computeASTHash}{$node$}
    \State $localHash \gets 0$
    \If{$isLeaf(node)$}
        \If{$isConstant(node)$}
            \State $localHash \gets combineHashes(node.sort.hashCode(), node.value.hashCode())$
        \Else
            \State $localHash \gets node.sort.hashCode()$ \Comment{$node$ is a variable}
        \EndIf
    \Else
        \State $localHash \gets combineHashes($
        \State $ \ \ \ \ node.sort.hashCode(), node.operator.hashCode(), node.numArguments.hashCode()$
        \State $)$
        \For{$child \in node.children$}
            \State $childHash \gets computeASTHash(child)$
            \State $localHash \gets combineHashes(localHash, childHash)$
        \EndFor
    \EndIf
    \State \Return $localHash$
\EndFunction
\end{algorithmic}
\caption{\added{Recursive function that computes hash for an AST node}}
\label{alg:hash_ast}
\end{algocolor}
\end{algorithm}

Algorithm for formula hash footprint computation is presented in~\ref{alg:hash_footprint}. Hash footprint is represented as a set that contains hash codes of all clauses of the formula. This yields a concise hash footprint that captures the structural properties of each clause in $F$.

\begin{algorithm}[tb]
\begin{algocolor}
\flushleft
\textbf{Input:} $F$ --- SMT formula\\
\textbf{Output:} $footprint$ --- hash footprint of the given SMT formula\\
\begin{algorithmic}[1]
\Function{computeFormulaHashFootprint}{$F$}
    \State $footprint \gets \emptyset$
    \For{$clause \in F.clauses$}
        \State $clauseHash \gets computeASTHash(clause)$
        \State $footprint \gets footprint \cup clauseHash$
    \EndFor
    \State \Return $footprint$
\EndFunction
\end{algorithmic}
\caption{\added{Hash footprint implementation}}
\label{alg:hash_footprint}
\end{algocolor}
\end{algorithm}

Although hash collisions can theoretically occur, using a robust Java hash function combined with an effective hash combination algorithm greatly reduces this risk. In practice, the hash footprint serves as an efficient initial check. If two formulae produce identical hash footprints, a more computationally intensive verification step is then performed to eliminate any false positives resulting from collisions (see Section~\ref{sec:testing}).

%% file: content/implementation.tex
\section{Implementation Details}
\label{section:impl}

The implementation\footnote{\url{https://github.com/plan-research/cache-a-lot}} of our approach, known as \cachealot{}, is written in Kotlin and built on top of the KSMT framework~\cite{KSMT}. KSMT provides essential data structures and an API facilitating interaction with various SMT solvers: Z3~\cite{z3}, Bitwuzla~\cite{bitwuzla}, CVC5~\cite{cvc5}, etc. Additionally, KSMT supports running multiple SMT solvers concurrently in the portfolio mode~\cite{portfolio}, leveraging Kotlin's coroutine-based concurrency model for efficient parallel operations~\cite{KotlinCoroutines}. These features allow \cachealot{} to operate independently of any specific solver, while still providing the advanced capabilities of KSMT.

\cachealot{} is designed to be used concurrently within Kotlin coroutines, allowing it to handle tasks like candidate selection and testing in a highly responsive and scalable manner. Furthermore, the implementation includes cancellation mechanisms, such as timeout-based cancellation of cache checks, providing greater control and flexibility during execution.

Another key aspect of \cachealot{} is its modularity. The implementation is structured to allow independent development and integration of different strategies for candidate selection and testing. This flexibility allows us to easily switch between different approaches, such as Utopia and our approach, by simply changing the testing strategy while keeping the storage and search mechanisms consistent. This design ensures that \cachealot{} can be adapted to various use cases and extended with new strategies.

%% file: content/evaluation.tex
\section{Evaluation}
\label{section:evaluation}

In this section, we present the design and setup of experiments conducted to evaluate the efficiency and effectiveness of \cachealot{}. We use a set of SMT formula benchmarks produced by symbolic and concolic program analysis tools. The primary goal is to determine whether our approach improves formula-solving performance through caching techniques and to compare it against the state-of-the-art approach, Utopia. Additionally, we aim to evaluate the effect of SMT formula canonization and the individual contributions of specific optimisations within \cachealot{}.

\subsection{Experimental Setup}

\begin{table}[tbh]
    \centering
    \scalebox{0.75}{
    \begin{tabular}{|c|c|c|c|c|c|c|c|c|}
         \hline
         Benchmark & \added{\multrow{Number of\\suites}} & \multrow{Number of\\formulae} & Sat (\%) & Unsat (\%) & Unknown (\%) & \multrow{Total\\solving time (s)} & \multrow{Avg.\\solving time\\for sat (ms)} & \multrow{Avg.\\solving time\\for unsat (ms)} \\
         \hline
         Klee & \added{99} & 181899 & 69.67 & 30.27 & 0.06 & \replaced{1101.20}{ 1100.33} & 2.62 & 3.58\\
         \hdashline
         Kex  & \added{89} & 11495 & 41.53 & 55.05 & 3.42 & \replaced{3267.63}{ 3272.86} & 176.69 & 66.39\\
         \hline
    \end{tabular}}
    \caption{Overview of the benchmarks used during evaluation}
    \label{tab:benchmarks}
\end{table}

\replaced{We evaluate our approach using two benchmarks:}{The evaluation is based on a set of 188 benchmarks, categorized into two primary groups:}
\begin{itemize}
    \item {\bf Klee benchmark\deleted{s}:} 99 \replaced{suites of SMT formulae}{ benchmarks} generated by Klee~\cite{Klee} during a symbolic analysis of 99 GNU Core Utilities\footnote{\url{https://www.gnu.org/software/coreutils/}} from the Linux kernel.
    
    \item {\bf Kex benchmark\deleted{s}:} 89 \replaced{suites of SMT formulae}{ benchmarks} derived from concolic analysis by Kex~\cite{kex}, covering 89 classes across 9 projects from the SBST 2022~\cite{sbst2022} and SBFT 2023~\cite{sbft2023} Java tool competitions.
\end{itemize}

We chose Klee and Kex due to their widespread use in symbolic and concolic testing, respectively. Klee produces simpler formulae with quantifier-free logic over bit-vectors and arrays~(QF\_ABV). At the same time, Kex also generates more complex formulae, including quantifiers, floating-point arithmetic, bit-vectors and arrays~(FPABV). This combination of benchmarks allows us to evaluate \cachealot{} across a wide range of SMT formula complexities.

In total, Klee generated 181,899 SMT formulae with an average solving time of 6 milliseconds per formula. Kex, on the other hand, produced 11,495 formulae with a significantly higher average solving time of 285 milliseconds. This difference in complexity makes Klee benchmarks suitable for testing \cachealot{}'s performance on relatively simple problems. In contrast, the Kex benchmarks push the system to handle more computationally demanding tasks. Table \ref{tab:benchmarks} provides a more detailed overview of the benchmarks.

We integrated Utopia's caching mechanism into our implementation for a fair and direct comparison between approaches. This ensures that both approaches are evaluated using the same solvers, benchmarks, and infrastructure, allowing us to isolate the effect of \cachealot{}'s design choices. \added{Additionally, our implementation removes some of the technical limitations of the original Utopia implementation~(e.g. support of quantifiers).}

Based on preliminary experiments, Bloom filter in the implementation was configured to use the fixed size $N = 1024$, both for our approach and Utopia. \added{We also explored different cache-emptying strategies. Our benchmarks, produced by symbolic execution tools, can be grouped by benchmark or suite. The experiments showed that suite-level caching is more efficient, as formulae within a single suite share more common components, unlike the more independent formulae at benchmark levels. Thus, we adopted suite-level caching in our evaluation, aligning with our goal to enhance the performance of symbolic execution-based tools using a local cache for each independent execution.}

The experiments were conducted using the KSMT framework~(v0.5.8) and Z3 solver~(v4.11.2). All experiments were run on a personal computer equipped with an 12th Gen Intel(R) Core(TM) i9-12900K processor, 64 GB of RAM, and Ubuntu 22.04.4 LTS.

\subsection{Research Questions}

Our experiments are designed to answer the following key research questions:

{\bf \rq{1}: Does \cachealot{} improve the efficiency of program analysis?}

This question evaluates whether \cachealot{} improves formula-solving efficiency compared to Utopia and vanilla solver. We focus on key metrics such as:
\begin{itemize}
    \item {\bf Time saved} is the reduction in formula-solving time achieved by reusing previously cached results. We calculate the time saved by comparing formula-solving time with and without caching mechanisms.
    \item {\bf Reuse ratio} measures the percentage of formulae that were successfully solved via a caching mechanism without actual invocation of the solver. It is a key indicator of how effective the caching system is at reusing previous results.
    \item {\bf Overhead} represents the total time spent on cache lookups during formula solving. This includes both successful cache hits and the overhead of unsuccessful cache lookups. The goal is to understand whether the potential savings from cache hits outweigh the costs of cache maintenance.
\end{itemize}

The comparison with Utopia is essential, as Utopia has demonstrated superior performance over other state-of-the-art solutions for caching SMT solver results. By comparing \cachealot{} directly with Utopia, we seek to determine whether \cachealot{} can offer meaningful improvements in terms of time saved and cache hit ratios. Additionally, we perform a qualitative analysis of the formulae that are successfully reused by each approach, examining whether the types of formulae differ significantly between the two methods. This comparison will allow us to assess quantitative metrics and the qualitative aspects of formula reuse.

{\bf \rq{2}: How does formula canonization affect the efficiency of the caching approaches?}

In this question, we investigate the effect of SMT formula canonization on caching performance. Canonization, which involves renaming variables in a consistent order, is used by Utopia to increase cache hit rates by making similar formulae more likely to match cached results.

\cachealot{} takes a different approach by not relying on canonization to mitigate the impact of variable renaming. This design decision means that \cachealot{} should be less affected by variations in variable names. In contrast, Utopia's performance may be affected by canonization, assuming improvement in effectiveness.

This experiment aims to clarify whether canonization offers a significant advantage for caching approaches like Utopia and how \cachealot{}'s resistance to such renaming strategies affects its overall performance.

\added{\protect\input{content/new/rq3}}

{\bf \rq{\replaced{4}{3}}: How do the optimisations improve the efficiency of the approach?}

\cachealot{} incorporates various optimisations designed to improve the efficiency of cache lookups and formula reuse. This question evaluates \deleted{the }individual contributions of each optimisation:
\begin{itemize}
    \item {\bf O1. Clause Hashing:} Grouping clauses by their hash codes for faster retrieval and reduced unification processes.
    \item {\bf O2. Invalid Substitution Filtering:} Filtering out invalid substitutions by intersecting variable sets before the join phase to minimize unnecessary computations.
    \item {\bf O3. Lazy Substitution Joining:} Using an iterator-based approach to process substitutions on-demand, reducing memory usage and avoiding the need to store large data tables.
\end{itemize}

We assess the impact of these optimisations by comparing \cachealot{} with and without them enabled, measuring their effect on cache hit rates and formula-solving times. This allows us to pinpoint which optimisations are most crucial for the performance gains.

\subsection{\rq{1} -- Efficiency}

\begin{figure}[tbh]
\begin{minipage}{0.3\linewidth}
    \centering
    \includegraphics[height=6.5cm]{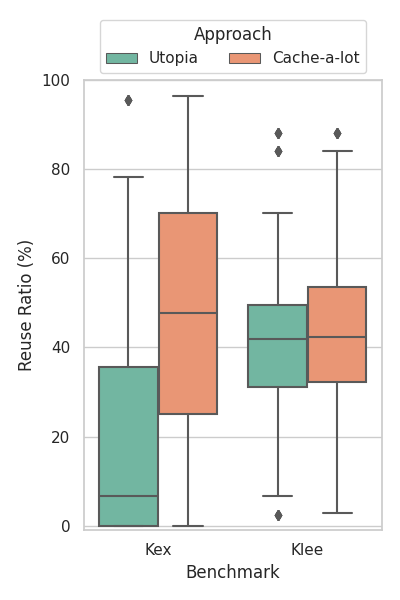}
    \caption{\centering Boxplot of Utopia and \cachealot{} reuse ratio's}
    \label{fig:reuse-rq1}   
\end{minipage}
\hspace{1em}
\begin{minipage}{0.6\linewidth}
    \centering
    \includegraphics[height=6.5cm]{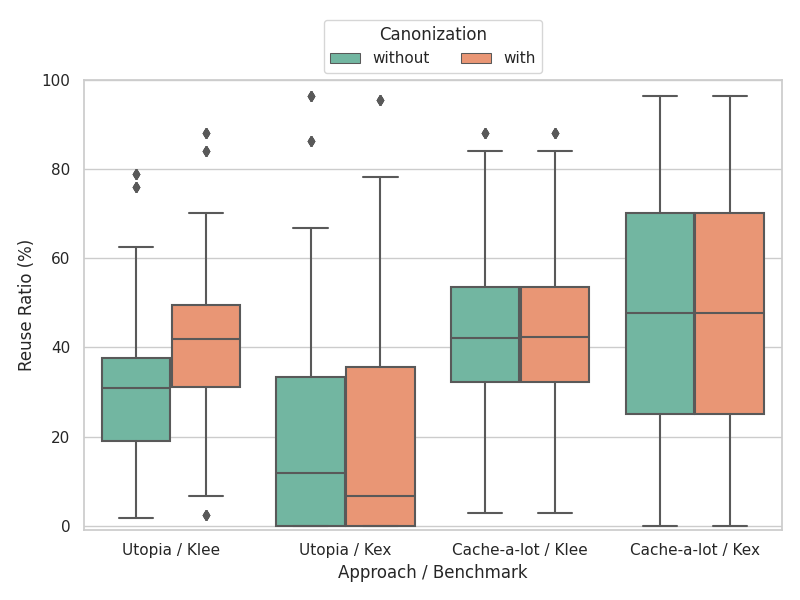}
    \caption{\centering Boxplot of Utopia and \cachealot{} reuse ratio's with and without formulae canonization}
    \label{fig:reuse-rq2}   
\end{minipage}
\end{figure}

\begin{table}[tbh]
    \centering
    \scalebox{0.9}{
    \begin{tabular}{|c|c|c|c|c|c|c|c|c|}
        \hline
        Benchmark & Tool & \multrow{\added{Overall}\\\added{solving}\\\added{time (s)}} & \multrow{\added{Unsat}\\\added{solving}\\\added{time (s)}} & \multrow{\added{Overall}\\\added{cache}\\overhead (s)} & \multrow{\added{Unsat}\\reuse\\ratio (\%)} & \multrow{Time\\saved \added{ratio}\\on unsat (\%)} & \multrow{\added{Overall}\\time\\saved (s)} \\
        \hline
        \added{Klee} & \added{No-cache} & \added{1101.20} & \added{206.79} & \added{0.00} & \added{0.00} & \added{0.00} & \added{0.00}\\
        \rowcolor{gray!10} Klee & \cachealot{} & \added{1045.76} & \added{151.26} & 0.27 & 34.47 & \replaced{26.85}{ 23.13} & 55.44 \\
        Klee & Utopia & \added{1049.01} & \added{154.47} & 0.41 & 33.30 & \replaced{25.30}{ 21.51} & 52.19\\
        \hdashline
        \added{Kex} & \added{No-cache} & \added{3267.63} & \added{441.22} & \added{0.00} & \added{0.00} & \added{0.00} & \added{0.00}\\
        \rowcolor{gray!10} Kex & \cachealot{} & \added{2940.14} & \added{111.09} & 7.87 & 74.36 & \replaced{74.82}{ 73.63} & 327.49\\
        Kex & Utopia & \added{3045.08} & \added{217.83} & 2.38 & 40.50 & \replaced{50.63}{ 53.08} & 222.55\\
        \hline
    \end{tabular}
    }
    \caption{Comparison of \cachealot{} and Utopia}
    \label{tab:t1_rq1}
\end{table}

In this section, we present the results of our experiments for evaluating \rq{1}, which focuses on measuring \cachealot{}'s efficiency improvements compared to Utopia. Both approaches were evaluated on the Klee and Kex benchmarks, each repeated multiple times to ensure consistency and minimize variability in the results. Each tool was executed 5 times on both benchmarks and we present averaged results.

Figure \ref{fig:reuse-rq1} shows the boxplot comparison of the reuse rate for \cachealot{} and Utopia on the Klee and Kex benchmarks, respectively. The reuse rate represents the percentage of formulae for which the caching mechanism successfully identified a previously solved result, reducing redundant computations. The figure illustrates that \cachealot{} consistently achieves a higher reuse rate on both benchmark sets, with a significant improvement on the more complex Kex benchmarks.

Table \ref{tab:t1_rq1} summarizes the averaged results across all executions, showing key metrics such as \replaced{solving time, solving time of unsat formulae, computational cost, unsat reuse ratio, and time saved}{ reuse ratio, total time saved, computational cost, and percentage of the time saved on unsat formulae}. \cachealot{} demonstrates a reuse ratio of 34.47\% on the Klee benchmarks, slightly outperforming Utopia's reuse ratio of 33.30\%. This improvement leads to \cachealot{} saving 55.44 seconds compared to Utopia's 52.19 seconds on the Klee benchmark. Although time saved is relatively small compared to the overall solving time of the benchmark~(1100.33 seconds),  the true impact of caching becomes more clear when we see that \cachealot{} saved \replaced{26.85\%}{ 23.13\%} of all time spent on unsat formulae compared to \replaced{25.30}{ 21.52\%} of Utopia. The overall structure of the Klee benchmarks that contains simpler formulae with the majority of sats makes it harder for caching to achieve a more noticeable result.
Additionally, we \deleted{can }see that both approaches incur minimal caching overhead on Klee, \cachealot{} maintains a slightly lower overhead of 0.27 seconds compared to Utopia's 0.41\deleted{ seconds}.

The difference between approaches becomes more pronounced on the Kex benchmarks, as it contains more complex formulae with a higher proportion of unsats. \cachealot{} achieves a reuse ratio of 74.36\%, significantly higher than Utopia's 40.50\%. This results in \cachealot{} saving 327.49 seconds, outperforming Utopia's 222.55 seconds. \cachealot{}'s advantage becomes even more obvious when comparing the relative times saved on unsat formulae: \replaced{74.82\%}{ 73.63\%} against \replaced{50.63\%}{ 53.08\%}. However, this improvement comes with a higher caching overhead of 7.87 seconds for \cachealot{}, compared to Utopia's 2.38 seconds. Despite the increased overhead, the time saved by \cachealot{} outweighs the additional cost, particularly for the Kex benchmarks, where formula complexity results in higher computation demands. Overall, we \deleted{can }conclude that \cachealot{} performs noticeably better than Utopia \deleted{on two sets of benchmarks} both in absolute~(overall time saved) and relative~(reuse ratio and time saved on unsat) metrics.

% Further breaking down the results into satisfiable and unsatisfiable formulae provides additional insight into the reuse efficiency of both approaches. For sat formulae, we cannot reduce solving time, so we expect overhead in such cases. On the other hand, for unsat formulae, we expect time savings. Table \ref{tab:t2_rq1} shows that both \cachealot{} and Utopia introduce minimal overhead on Sat formulae, which means no significant performance reduction in such cases. Regarding the Kex benchmark, \cachealot{} introduces a negligible 0.06\% overhead, slightly higher than Utopia's 0.04\%.

% However, the true impact of caching emerges when analyzing the results for Unsat formulae. \cachealot{} demonstrates much higher time savings on Unsat formulae across both benchmark sets. On the Kex benchmarks, \cachealot{} saves 66.33\% of the total solving time, compared to Utopia's 25.79\%. Even on the simpler Klee benchmarks, \cachealot{}'s 25.14\% time savings exceed Utopia's 17.13\%.

\begin{figure}[tbh]
\begin{subfigure}{0.3\linewidth}
    \centering
    \includegraphics[width=\linewidth]{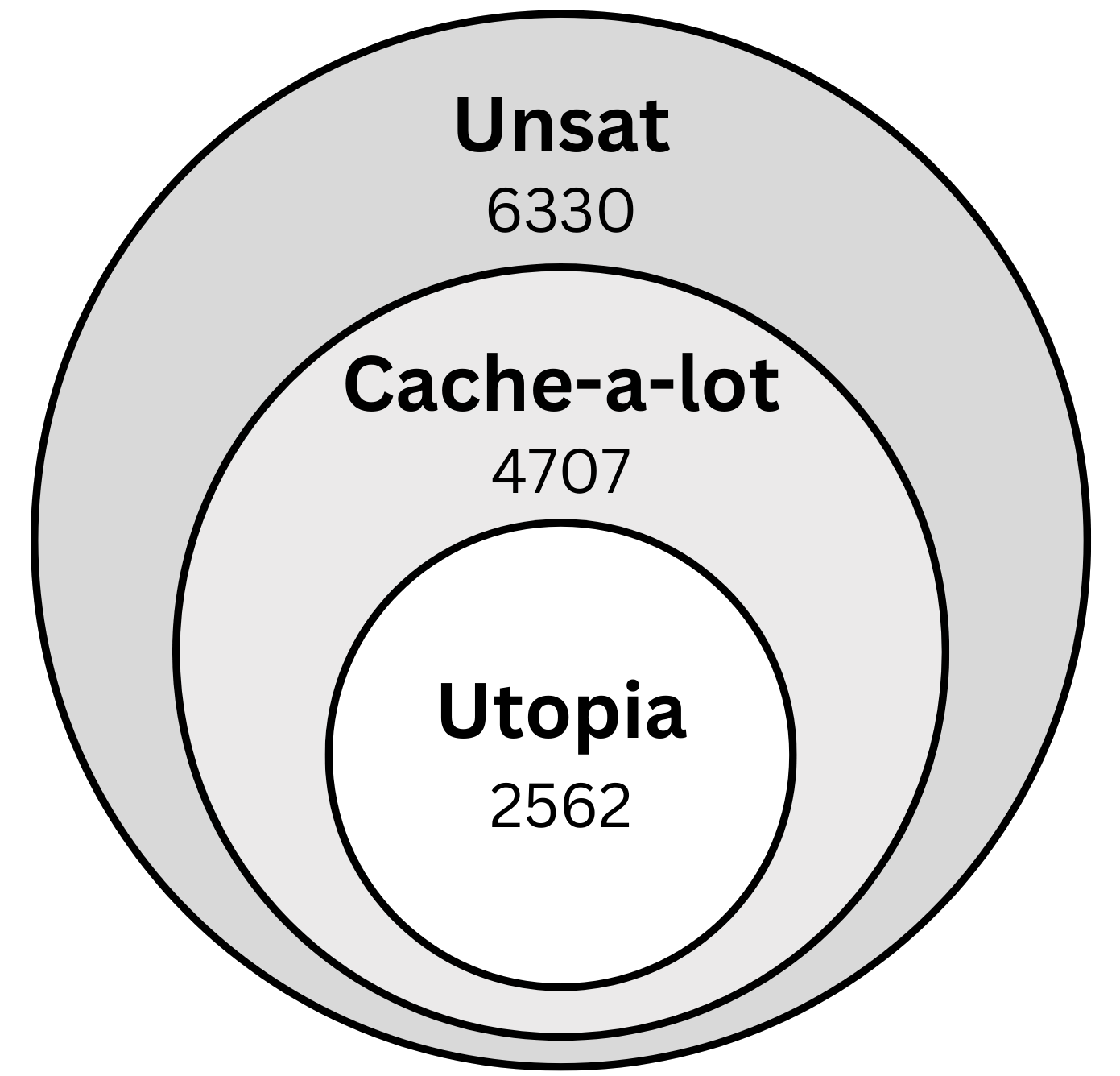}
    \caption{Kex}
    \label{fig:circ_kex}
\end{subfigure}\hspace{0.1\linewidth}
\begin{subfigure}{0.3\linewidth}
    \centering
    \includegraphics[width=\linewidth]{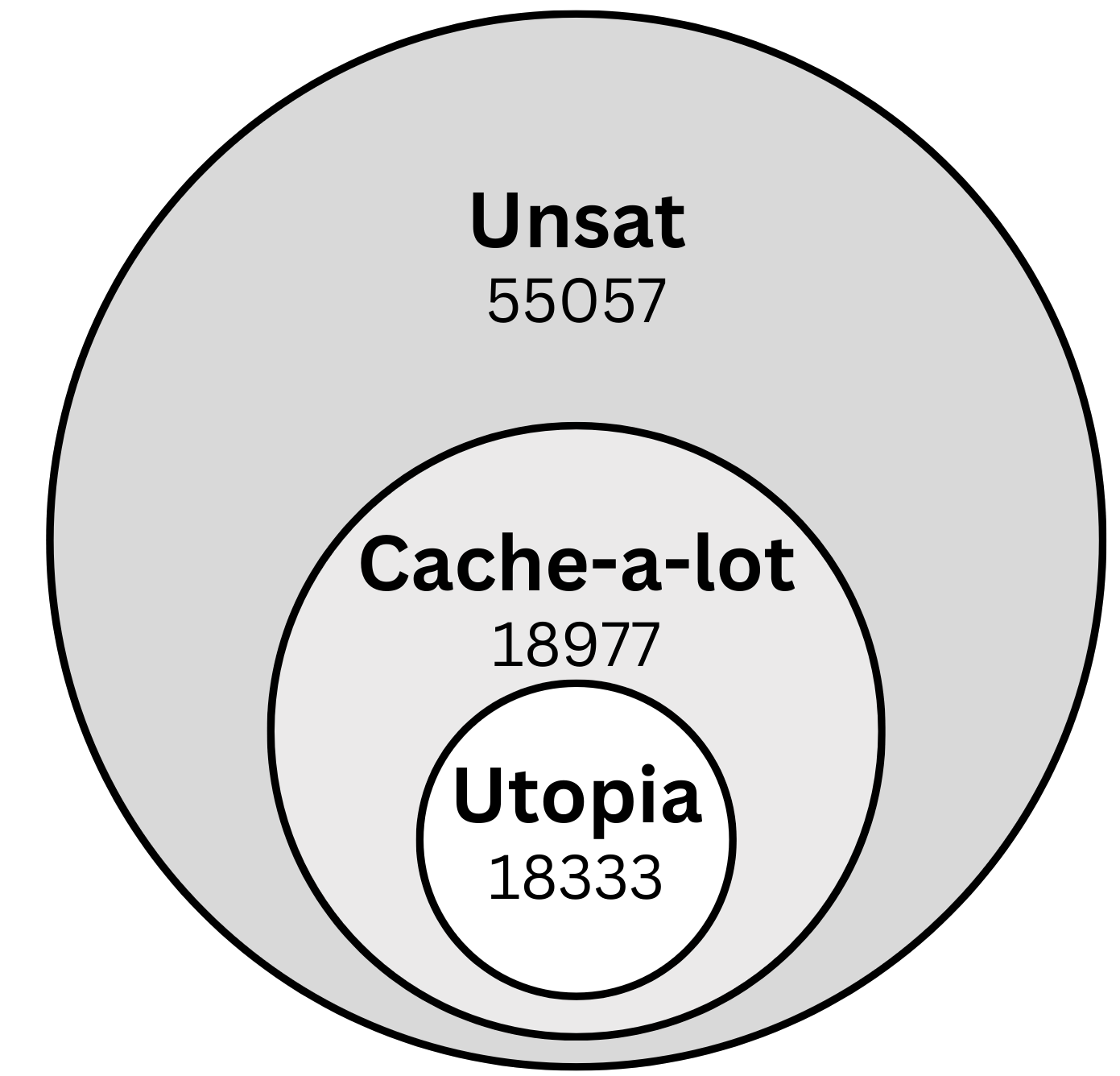}
    \caption{Klee}
    \label{fig:circ_klee}
\end{subfigure}
\caption{Venn diagram of unsat reusage}
\label{fig:circ}
\end{figure}

In addition to the quantitative results, a qualitative analysis of the reused formula sets reveals further insights into \cachealot{}'s superior efficiency~(figure~\ref{fig:circ}). The set of formulae reused by Utopia is consistently a subset of those reused by \cachealot{}, both for Kex~(figure~\ref{fig:circ_kex}) and Klee~(figure~\ref{fig:circ_klee}) benchmarks, highlighting \cachealot{}'s broader coverage. This means that \cachealot{} not only captures all formulae that Utopia would reuse but also manages to reuse additional formulae that Utopia misses, further increasing the overall reuse rate.

Overall, these results demonstrate that \cachealot{} offers significant improvements in formula-solving efficiency, particularly for complex formulae \replaced{of}{such as those found in} the Kex benchmark. Despite the additional computational overhead, the broader reuse of formulae leads to valuable time savings and a higher reuse rate, resulting in superior performance both quantitatively and qualitatively.

\subsection{\rq{2} -- Canonization Effect}

In this section, we evaluate how SMT formula canonization affects caching performance for both \cachealot{} and Utopia. Canonization, as implemented in Utopia, involves renaming variables in a standardized order to increase cache hit rates by making formulae with similar structures more likely to match cached results. \cachealot{}, however, does not rely on canonization, as its caching mechanism is designed to be more robust to variable renaming and thus should not benefit from this technique.

The boxplots in Figure \ref{fig:reuse-rq2} provide a visual comparison of the reuse ratios for Utopia and \cachealot{} with and without SMT formula canonization across different benchmarks. From the data, it becomes evident that Utopia is significantly affected by canonization, with its performance varying depending on the benchmark.

On the Klee benchmark, Utopia shows a higher reuse ratio when using canonization. Specifically, Utopia achieves a reuse ratio of 33.30\% and saves 52.19 seconds of solving time with canonization. Without canonization, \deleted{the }reuse ratio drops to 24.29\%, with 40.32 seconds of saved time. This indicates that canonization substantially improves Utopia's ability to reuse formulae on the simpler Klee benchmark, where standardized renaming helps in matching previously cached results.

On the Kex benchmarks, the effect of canonization is more nuanced. Utopia achieves a slightly higher reuse ratio without canonization, 41.23\% compared to 40.50\% with canonization. However, despite the lower reuse ratio with canonization, Utopia still saves significantly more time: 222.55 seconds compared to 105.25 seconds without canonization. This suggests that, while canonization leads to fewer reuse opportunities, the formulae that are reused contribute to more substantial time savings. Thus, Utopia with and without canonization appears to reuse qualitatively different formulae, with canonization enabling the reuse of more computationally expensive formulae.

In contrast, the boxplots show that \cachealot{} is unaffected by canonization. Its reuse ratio remains consistent across both Klee and Kex benchmarks, regardless of whether canonization is applied. This is expected, as \cachealot{}'s caching mechanism is designed to be robust to variations in variable naming, meaning that it can reuse formulae effectively without relying on canonization. As a result, there is no observable difference in its performance with or without canonization, further highlighting the robustness of \cachealot{}'s approach.

Overall, the results indicate that canonization significantly influences Utopia's performance, with its effectiveness varying depending on the benchmark and formula characteristics. On the simpler Klee benchmarks, canonization clearly improves both the reuse ratio and time saved. However, on the more complex Kex benchmarks, the reuse ratio is slightly better without canonization, but the time saved is significantly higher with it, suggesting that canonization enables the reuse of more computationally expensive formulae.

\cachealot{}, on the other hand, remains unaffected by canonization, offering consistent performance across benchmarks. This demonstrates that \cachealot{}'s caching mechanism is inherently robust to variable renaming, providing a more flexible and adaptable approach to formula reuse compared to Utopia.

\added{\protect\input{content/new/rq3-expr}}

\subsection{\rq{\replaced{4}{3}} -- Optimisations Effectiveness}

In this section, we assess the individual contributions of various optimisations in \cachealot{} and their combined effects on formula-solving performance. To measure the effects of these optimisations, we executed six versions of \cachealot{}:
\begin{itemize}
    \item Without any optimisations.
    \item With only O1.
    \item With only O2.
    \item With only O3.
    \item With both O2 and O3 combined.
    \item With all optimisations enabled~(O1+O2+O3).
\end{itemize}

O1 optimisation is independent of O2 and O3 and does not affect them. As a result, there is no need to evaluate combinations of O1 with the other optimisations. O2 and O3, however, can influence each other, making it essential to examine their combined effect.

Each version was executed 5 times on the Kex benchmark suite. The Kex benchmarks were selected because they contain more complex formulae, meaning any performance differences will be even more pronounced. \replaced{We executed each version of \cachealot{} with five different heap size configurations~(768\,MB, 1\,GB, 2\,GB, 4\,GB, and 8\,GB) and measured the overall lookup overhead.}{However, the base version~(without optimisations) and the version with O1 failed to complete execution on 6 out of the 89 Kex benchmarks due to out-of-memory errors. For consistency, these benchmarks were excluded from the final analysis.  We measured only time spent on the cache execution, as it is the only metric that will be affected by optimisations.}

\ifdraft
\begin{table}[tbh]
    \centering
    \begin{tabular}{|c|c|c|}
         \hline
         \deleted{Version} & \deleted{Execution time (s)} & \deleted{Time reduction (\%)} \\
         \hline
         \rowcolor{gray!10} \deleted{Base} & \deleted{19.62} & \deleted{--}\\
         \deleted{O1} & \deleted{13.55} & \deleted{-30.93}\\
         \rowcolor{gray!10} \deleted{O2} & \deleted{5.70} & \deleted{-70.94}\\
         \deleted{O3} & \deleted{12.88} & \deleted{-34.35} \\
         \rowcolor{gray!10} \deleted{O2 + O3} & \deleted{5.63} & \deleted{-71.30}\\
         \deleted{O1 + O2 + O3} & \deleted{1.55} & \deleted{-92.10} \\
         \hline
    \end{tabular}
    \caption{\deleted{Execution time of \cachealot{} with different optimisations}}
    \label{tab:opt}
\end{table}
\fi

% \begin{table}[tbh]
%     \centering
%     \begin{tabular}{|c|c|c|}
%          \hline
%          Version & Execution time (s) & Time reduction (\%) \\
%          \hline
%          \rowcolor{gray!10} Base & 19.62 & --\\
%          O1 & 13.55 & -30.93\\
%          \rowcolor{gray!10} O2 & 5.70 & -70.94\\
%          O3 & 12.88 & -34.35 \\
%          \rowcolor{gray!10} O2 + O3 & 5.63 & -71.30\\
%          O1 + O2 + O3 & 1.55 & -92.10 \\
%          \hline
%     \end{tabular}
%     \caption{Execution time of \cachealot{} with different optimisations}
%     \label{tab:opt}
% \end{table}

\begin{table}[tbh]
\begin{minipage}{0.35\linewidth}
\centering
    \begin{tabular}{|c|c|}
         \hline
         \added{Version} & \multrow{\added{Lookup}\\\added{overhead (s)}} \\
         \hline
         \rowcolor{gray!10} \added{Base} & \added{---} \\
         \added{O1} & \added{---} \\
         \rowcolor{gray!10} \added{O2} & \added{35.19} \\
         \added{O3} & \added{64.61} \\
         \rowcolor{gray!10} \added{O2 + O3} & \added{33.66} \\
         \added{O1 + O2 + O3} & \added{8.86} \\
         \hline
    \end{tabular}
    \caption{\added{Lookup overhead time of \cachealot{} with different optimisations on 8\,GB heap}}
    \label{tab:opt}
\end{minipage}\hfill
\begin{minipage}{0.60\linewidth}
\centering
    \centering
    \includegraphics[width=\linewidth]{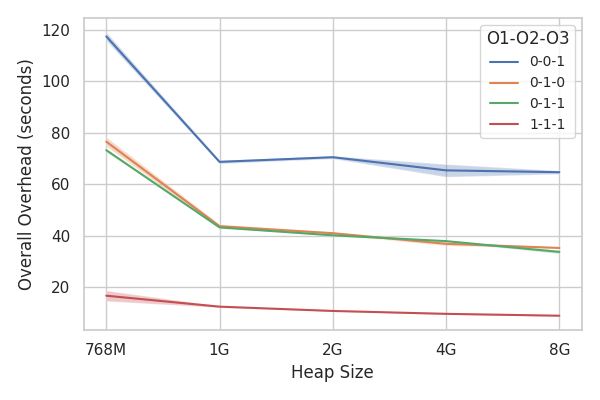}
    \captionof{figure}{\centering \added{Dependence between Allocated Heap Memory and Lookup Overhead for \cachealot{} optimisations.}}
    \label{fig:opt} 
\end{minipage}
\end{table}

\replaced{
Table~\ref{tab:opt} shows the overall lookup overhead for each of the \cachealot{} versions on 8\,GB heap, while figure~\ref{fig:opt} shows the dependence between allocated heap memory and overall lookup overhead for each version. First of all we can see, that base~(without optimisations) and O1 versions did not finish execution on any of the benchmarks due to out-of-memory errors. Second, we can see that both O2 and O3 optimisations successfully finished their execution. This means that O2 and O3 not only optimise the performance of the \cachealot{}, but also reduce its memory consumption.  Combination of all the optimisations~(O1+O2+O3) performs best, and, most notably, better than any individual version.}{The total averaged time consumption for each version on the filtered Kex benchmarks is summarized in Table \ref{tab:opt}. These results demonstrate that each optimisation has a significant impact of optimisations on the approach's performance. O2 has shown the greatest single impact on the overall execution time. Moreover, while the main purpose of O3 is optimising the memory usage, we can see that it positively affects performance both single handedly and in combination with other optimisations.}

%% file: content/new/rq3.tex
{\bf \rq{3}: What is the impact of caching mechanisms on the memory consumption?}

This question examines three key aspects of memory usage in caching for SMT solving:
\begin{itemize}
    \item {\bf Static Memory Consumption:} What is the amount of memory required to maintain the cache data structures, and how does it influence overall system resource usage?

    \item {\bf Peak Memory Consumption during Cache Lookups:} How much memory is consumed at the peak during cache lookup operations, such as unification and substitution joining, and what are the implications for scalability?

    \item {\bf Dependence between Allocated Heap Memory and Lookup Overhead:} How does the variation in allocated heap memory correlate with lookup time/overhead?
\end{itemize}

By comparing these memory-performance trade-offs between \cachealot{} and Utopia, we aim to determine whether the benefits of enhanced caching efficiency justify the associated memory overhead.

%% file: content/new/rq3-expr.tex
\subsection{\rq{3} -- Memory Consumption Impact}

In this section, we present the experimental evaluation of \rq{3}, which investigates the impact of memory consumption on the efficiency of caching mechanisms.

\subsubsection{Static Memory Consumption.}
\hfill

To evaluate the static memory footprint of the caching mechanisms, we measured memory consumption immediately after solving each formula. Specifically, after processing a formula, we explicitly invoked the garbage collector using \texttt{System.gc()} and then recorded the memory usage via the \texttt{Runtime} object. We obtained a sequence of memory consumption values for each benchmark suite and computed the maximum static memory consumption for it. Then, we performed a paired t-test on these maximum values to assess whether there is a statistically significant difference between \cachealot{} and Utopia.

\begin{figure}[tbh]
\begin{minipage}{0.45\linewidth}
    \centering
    \includegraphics[width=0.8\linewidth]{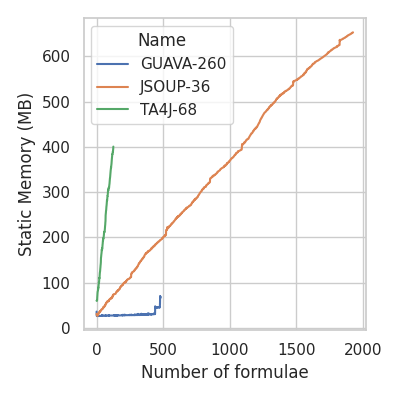}
    \caption{\added{Examples of static memory consumption per formula for \cachealot{} and Utopia.}}
    \label{fig:static_memory_charts}
\end{minipage}
\hspace{1em}
\begin{minipage}{0.45\linewidth}
    \centering
    \includegraphics[width=0.9\linewidth]{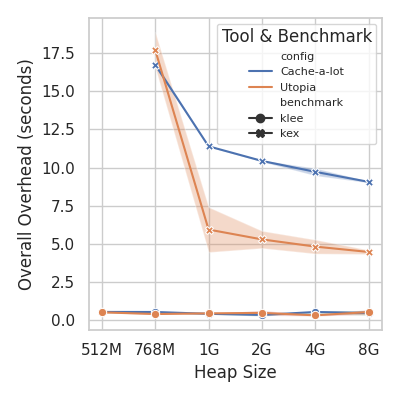}
    \caption{\added{Dependence between Allocated Heap Memory and Lookup Overhead for \cachealot{} and Utopia on Klee and Kex. Error bands represent variability across multiple runs.}}
    \label{fig:heap_lookup_overhead} 
\end{minipage}
\end{figure}

% \begin{figure}[tbh]
%     \centering
%     \includegraphics[width=0.8\linewidth]{content/new/rq3-examples.png}
%     \caption{Examples of static memory consumption per formula for \cachealot{} and Utopia.}
%     \label{fig:static_memory_charts}
% \end{figure}

Static memory consumption of the external cache is influenced by several factors, including the number of unsatisfiable formulas processed, the reuse ratio of the caching mechanism, and the size of the unsat cores or formulas. Our experiments reveal distinct behaviors in the evolution of static memory consumption per formula. Figure~\ref{fig:static_memory_charts} shows the growth of static memory consumption depending on the number of processed formulae for several of our suites.

Our results show that for the Klee benchmarks, \cachealot{} recorded a maximum static memory consumption of 32.82\,MB compared to 32.69\,MB for Utopia, with a p-value of 0.43. Similarly, for the Kex benchmarks, the maximum static memory consumption was 652.86\,MB for \cachealot{} and 652.82\,MB for Utopia, with a p-value of 0.37. These high p-values indicate that the differences in maximum static memory consumption between the two approaches are not statistically significant. Thus we can conclude, that both \cachealot{} and Utopia have statistically similar static memory consumption.

\subsubsection{Peak Memory Consumption during Cache Lookups.}
\hfill

To assess the peak memory consumption during cache lookup operations, we continuously measured memory usage every 50\,ms during each lookup phase. For each formula, we computed the peak memory consumption as the maximum memory observed during lookup minus the static memory measured immediately after formula solving. Then, for each benchmark suite, we obtained the maximum of these per-formula measurements and averaged them among $3$ executions. We also computed the p-values over these maximum values for statistical comparison. Note that during lookup, the garbage collector was not called manually, and the available heap space was set to 15\,GB.

\begin{table}[tbh]
    \centering
    \scalebox{1.0}{
    \begin{tabular}{|c|c|c|c|}
        \hline
        \added{Benchmark} & \added{Tool} & \multrow{\added{Peak Lookup}\\\added{Memory (MB)}} & \added{p-value} \\
        \hline
        \added{Klee} & \added{\cachealot{}} & \added{3.38} & \multirow{2}{*}{\added{0.35}} \\
        \added{Klee} & \added{Utopia}     & \added{4.20} & \\
        \hdashline
        \added{Kex}  & \added{\cachealot{}} & \added{354.02} & \multirow{2}{*}{\added{0.085}} \\
        \added{Kex}  & \added{Utopia}     & \added{45.31}  & \\
        \hline
    \end{tabular}
    }
    \caption{\added{Summary of maximum peak lookup memory consumption for \cachealot{} and Utopia on the Klee and Kex benchmarks.}}
    \label{tab:peak_memory}
\end{table}

Table~\ref{tab:peak_memory} reports the maximum peak lookup memory consumption for each benchmark and for both \cachealot{} and Utopia, along with the corresponding p-values. The results confirm that, as expected, \cachealot{} consumes more memory during lookup operations on the more complex Kex benchmarks. In contrast, the two approaches exhibit similar peak memory profiles on the simpler Klee benchmarks. The p-values indicate that the difference in peak lookup memory is statistically significant for Kex (p = 0.085, approaching significance) but not for Klee (p = 0.35).

% Table~\ref{tab:peak_memory} reports the maximum peak lookup memory consumption for each benchmark and for both \cachealot{} and Utopia, along with the corresponding p-values. For the Kex benchmarks, Utopia showed a maximum peak lookup memory of 45.31\,MB, whereas \cachealot{} reached 354.02\,MB on average. In contrast, for the Klee benchmarks, Utopia and \cachealot{} exhibited maximum peak lookup memories of 4.20\,MB and 3.38\,MB, respectively. The p-values indicate that the difference in peak lookup memory is statistically significant for Kex (p = 0.085, approaching significance) but not for Klee (p = 0.35).

% These results confirm that, as expected, \cachealot{} consumes more memory during lookup operations on the more complex Kex benchmarks. In contrast, the two approaches exhibit similar peak memory profiles on the simpler Klee benchmarks.

\subsubsection{Dependence between Available Heap Memory and Lookup Overhead.}
\hfill

We further explored the relationship between available heap space and the overhead incurred during cache lookups by running both \cachealot{} and Utopia under six different heap size configurations: 512\,MB, 768\,MB, 1\,GB, 2\,GB, 4\,GB, and 8\,GB. For each configuration, we measured the total (summed) lookup overhead across all formulas in the Klee and Kex benchmarks.

% \begin{figure}[tbh]
%     \centering
%     \includegraphics[width=0.8\linewidth]{content/new/heap_overhead.png}
%     \caption{Dependence between Allocated Heap Memory and Lookup Overhead for \cachealot{} and Utopia on Klee and Kex. Error bands represent variability across multiple runs.}
%     \label{fig:heap_lookup_overhead}
% \end{figure}

As shown in Figure~\ref{fig:heap_lookup_overhead}, both tools experienced \emph{out-of-memory} errors on the more complex Kex benchmark at the lowest heap setting (512\,MB), since the static memory usage alone exceeded this limit. Beyond this point, increasing the heap size consistently led to lower lookup overhead, highlighting a clear trade-off between memory allocation and cache performance. In other words, while larger heap allocations help avoid memory pressure and reduce overhead, they also represent higher resource usage. For the simpler Klee benchmark, the lookup overhead remained comparatively small under all tested heap sizes, indicating that more substantial heap space primarily benefits complex analyses such as those in Kex.

%% file: content/new/discussion.tex
\section{Discussion}
\label{section:discussion}

In this section, we situate our unsatisfiable core caching methodology within the broader landscape of SMT solving techniques. We compare our approach against established paradigms --- namely, graph-theoretic algorithms, incremental solving, and knowledge compilation --- highlighting both the conceptual synergies and practical trade-offs. This discussion not only underscores the theoretical foundations of our work but also outlines promising directions for future research.

\textbf{Graph Algorithms.} Graph-based methods offer a promising way to capture structural similarities in SMT formulas by representing them as graphs. However, we found that current algorithms are not well suited for simultaneously detecting isomorphisms among multiple subgraphs, a requirement crucial for our application. This challenge, rather than a limitation of our approach, reflects the inherent complexity of the problem and motivates further research in efficient multi-subgraph isomorphism detection.

\textbf{Incremental Solving.} Our current work is based on traditional batch-mode SMT solving, where each formula is handled independently. In contrast, incremental SMT solvers use \textit{push/pop} operations to modify formulas incrementally. It allows to implement methods designed to take advantage of similarities between formulas in a dynamic setting. Our unsat core caching mechanism could be adapted to work with incremental solvers by aligning cache updates with these operations. Moreover, it can be implemented more efficiently. For example, we can keep the already computed unifying substitutions while the formula's clause is in the formula.

\textbf{Knowledge Compilation.} Knowledge compilation transforms logical formulas into alternative representations to enable more efficient reasoning. While this approach can potentially precompile unsat cores and facilitate rapid reuse, it also has certain drawbacks. In particular, the additional memory overhead and preprocessing time required to generate these representations can counterbalance the benefits.

%% file: content/conclusion.tex
\section{Conclusion}
\label{section:conclusion}

In this paper we present \cachealot{}, a novel approach that pushes the limits of unsatisfiable core reuse in SMT-based program analysis. Unlike existing state-of-the-art approaches, \cachealot{} does not rely on formula canonization and instead considers all the potential variable substitutions during unsat core selection and testing. This allows \cachealot{} to reuse great proportion~(34\%--74\%) of all the unsat cores during the analysis, all while being efficient enough to ensure significant time saves as well.

Our experimental results show that our approach outperforms state-of-the-art Utopia approach both in terms of the percentage of reused unsat cores and in terms of overall time saved due to caching. Additionally, we demonstrate that our approach does not rely on any type of formula preprocessing, while the performance of existing approaches can vary heavily depending on the quality of preprocessing. Moreover, we demonstrate that candidate testing optimizations, introduced by \cachealot{} have a great impact on the overall efficiency of our approach. 

Current solver-agnostic implementation of \cachealot{}, build on top of KSMT framework, allows it to be easily adapted and expanded, opening a lot of potential for the future development. Our future plans include developing an SMT model reuse approach that will allow us to impact satisfiable SMT formulae as well. Additionally, we plan to adapt \cachealot{} to support incremental SMT formulae as well, as these types of formulae have a lot of potential for SMT solutions reuse.

\section{Data availability}

Reproduction package with the source code of our implementation, both benchmarks and the results presented in the evaluation is available at~\replaced{\url{https://doi.org/10.5281/zenodo.14925823}}{\\\url{https://doi.org/10.5281/zenodo.13750120}}.